\RequirePackage[2020-02-02]{latexrelease}
\documentclass[twocolumn, switch]{article} 

\usepackage{preprintvr}

\usepackage{amsmath, amsthm, amssymb, amsfonts}

\usepackage[numbers,square]{natbib}
\usepackage[utf8]{inputenc} 
\usepackage[T1]{fontenc}  
\usepackage{xcolor}   
\usepackage[colorlinks = true,
            linkcolor = purple,
            urlcolor  = blue,
            citecolor = cyan,
            anchorcolor = black]{hyperref}  
\usepackage{booktabs}     
\usepackage{nicefrac}   
\usepackage{microtype}    
\usepackage{lineno}   
\usepackage{float}      

\usepackage{lipsum}   

\usepackage{newfloat}
\DeclareFloatingEnvironment[name={Supplementary Figure}]{suppfigure}
\usepackage{sidecap}
\sidecaptionvpos{figure}{c}

\usepackage{titlesec}
\titlespacing\section{0pt}{12pt plus 3pt minus 3pt}{1pt plus 1pt minus 1pt}
\titlespacing\subsection{0pt}{10pt plus 3pt minus 3pt}{1pt plus 1pt minus 1pt}
\titlespacing\subsubsection{0pt}{8pt plus 3pt minus 3pt}{1pt plus 1pt minus 1pt}

\usepackage{tikz,xcolor,hyperref}

\definecolor{lime}{HTML}{A6CE39}
\DeclareRobustCommand{\orcidicon}{
  \begin{tikzpicture}
  \draw[lime, fill=lime] (0,0) 
  circle [radius=0.16] 
  node[white] {{\fontfamily{qag}\selectfont \tiny ID}};
  \draw[white, fill=white] (-0.0625,0.095) 
  circle [radius=0.007];
  \end{tikzpicture}
  \hspace{-2mm}
}
\foreach \x in {A, ..., Z}{\expandafter\xdef\csname orcid\x\endcsname{\noexpand\href{https://orcid.org/\csname orcidauthor\x\endcsname}
      {\noexpand\orcidicon}}
}

\title{Privacy concerns from variances in spatial navigability in VR}

\usepackage{xwatermark}
\newwatermark[firstpage,color=gray!60,angle=90,scale=0.32, xpos=3.9in,ypos=0]{\href{https://doi.org/}{\color{gray}{Preprint doi}}}
\newwatermark[firstpage,color=gray!90,angle=0,scale=0.28, xpos=0in,ypos=-5in]{*correspondence: \texttt{abasu@ualr.edu}}

\usepackage{authblk}

\author[1]{Aryabrata Basu\orcidA{}}
\author[1]{Mohammad Jahed Murad Sunny\orcidB{}}
\author[1]{Jayasri Sai Nikitha Guthula\orcidC{}}

\affil[1]{Department of Computer Science, University of Arkansas at Little Rock}

\begin{document}

\twocolumn[ 
  \begin{@twocolumnfalse} 
  
\maketitle

\begin{abstract}
Current Virtual Reality (VR) input devices make it possible to navigate a virtual environment and record immersive, personalized data regarding the user's movement and specific behavioral habits, which brings the question of the user's privacy concern to the forefront. In this article, the authors propose to investigate Machine Learning driven learning algorithms that try to learn with human users co-operatively and can be used to countermand existing privacy concerns in VR but could also be extended to Augmented Reality (AR) platforms. 
\end{abstract}
\vspace{0.35cm}

  \end{@twocolumnfalse} 
] 



\section{Introduction}
Navigating spaces is an embodied experience. Spatial navigation in a VR environment requires auditory, visual, and or tactile sensation to explore the virtual space intertwined with tasks to perform. VR devices expand the dimension of interaction points in that environment. The latest motion capture device, like Sony's Mocopi \cite{mocopi}, can capture and operate a 3d avatar in a VR environment using real-world physical movements of human limbs and tracking and translating them in real-time with a minimal setup footprint. The state of current VR input devices' tracking implications is immense and can reduce the barrier to longterm VR usability. At the same time, people are concerned about these tracked datasets, which run the risk of violating user privacy in different ways of inferencing \cite{o2016convergence}\cite{10.1145/2580723.2580730}; physical harms \cite{Cobb1999VirtualRS}, and manipulation and violation of immersive experiences \cite{10.5555/3291228.3291263} \cite{Kerr2008CriminalLI}. 

In the event of a VR system security flaw, it is possible to exploit VR usability data to produce malware for users, putting them at serious risk \cite{Safetyandprivacy}. Imposing the control point to the data flow through VR technology has become vital as the technology space has evolved at a breakneck rate.  While engaging in various interesting VR activities, a well-known platform called Horizon Worlds by Meta actively records user behavioral data, including audio logs, as part of their data retention policy. However, it is known as `doxing' when information from social media networking sites is exploited to reveal users' private lives. One may utilize the Metaverse to gain access to more private information about a person, like their habits and physical traits \cite{kurtunluouglu2022security}.
The sections [~\ref{sec:testcase}] and [~\ref{sec:measure}] of this article try to address with a demonstration why the significance of security in VR should be elevated.

\section{Background}

Access to personalized information about a user can be used against them to alter how they interact with any information resource. Typically, the information saved about a user is referred to as a user model or profile. An established area of artificial intelligence research is user modeling. Many adaptive or personalized systems have an embedded user model that is the basis for features like customized recommendations. This strategy, which has several limitations, has been noted to require the application developer to create and maintain their own user model. Because of these limitations, Fortnite  \cite{Fortnite}, a popular video game, took advantage of a privacy loophole and used digital patterns to draw information from kids under 13 years old without obtaining the consent of their parents. The parent company, Epic Games, went through a settlement with Federal Trade Commission for a privacy violation\cite{FortniteArticle}. 

According to Kanoje et al., user profiling is the practice of locating information about a user's realm of interest. The system can use this data to learn more about the user and utilize this understanding to improve retrieval so that the user is satisfied \cite{kanoje2015user}. In another article on user profiling by Farid et al. \cite{farid2018user}, the creation of an initial user profile for a new user and the ongoing updating of the profile information to accommodate the ongoing changes in the user's interests, preferences, and needs are the two ways used in the user modeling process. Creating a user profile is a significant difficulty in user profiling to represent user preferences correctly. But this data is being studied even though privacy is violated. So to learn more about user profiling, we considered a test case to identify a few of the hidden dimensions potentially exploitative in nature.

\begin{table}[H]
  \begin{center}
    \begin{tabular}{@{}p{5cm}p{3cm}@{}}\toprule
      Features & Definition \\ \toprule
      Distance traveled & It is the total length of the path traveled between two positions. \\ \midrule
      Coverage & Total number of unit cubes covered (area). \\ \midrule
      Number of decision points reached & Total number of decision points covered in the maze.  \\ \midrule
      Positional Curvature & The signed angle of curvature between two consecutive position vectors.   \\ \midrule
      Head rotation amount & The unsigned head rotation angle between two consecutive rotation transforms.  \\ 
      \bottomrule
    \end{tabular}
    
  \end{center}
  \caption [User trajectory features]
  {Features extracted from users' trajectory data for deep exploration and their corresponding definitions.}
  \label{table:ucavecurrentmazefeaturestable}
\end{table}

\section{Test case scenario : A single-level maze}
\label{sec:testcase}
\begin{figure}[th]
 \centering 
 \includegraphics[width=0.8\linewidth]{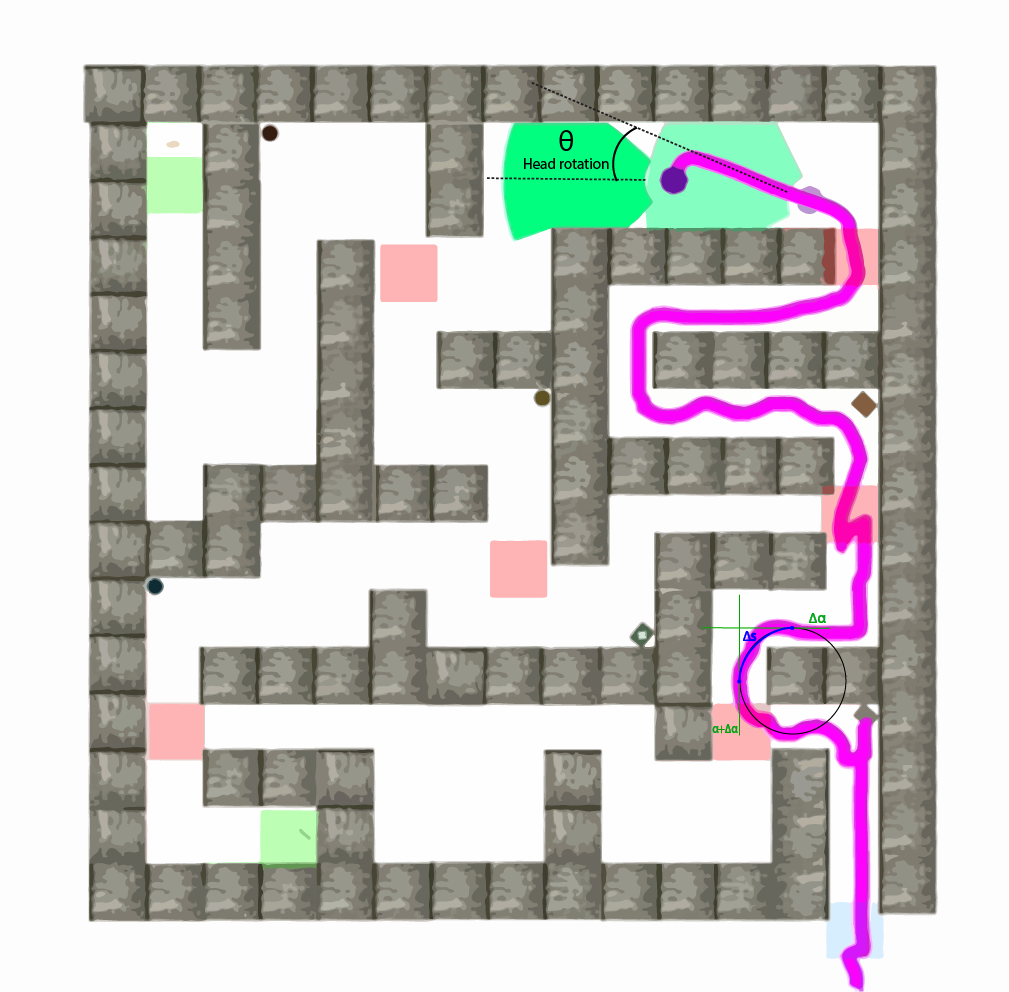}
 \caption [Single level maze study - Visualization of two user trajectory features]
 {Single level maze study - Visualization of two trajectory features: positional curvature and head rotation amount.}
 \label{fig:featurestrajectory}
\end{figure}

A simple (single level) 3d maze was designed to observe navigability of the subjects while a set of empirical data were collected such as, trajectory data, head rotation, and body rotation [Figure \ref{fig:featurestrajectory}]. The maze experience was deployed as a 2×2 study design. A 2×2 factorial design is an experimental design that allows researchers to understand the effects of two independent variables (each with two levels) on a single dependent variable.

\section{Measures}
\label{sec:measure}
To conduct a thorough analysis of the trajectory data, we defined a set of mathematically derived features that were generalized for each trajectory. These features included the distance traveled, the number of decision points (nodes) reached inside the maze, the amount of head rotation, the positional curvature, and the maze coverage. Some of these trajectory features are illustrated in Table \ref{table:ucavecurrentmazefeaturestable}. Positional curvature feature refers to the curvature of the trajectory calculated per frame between successive position vectors. Rotation amount feature refers to the unsigned angle calculated per frame between successive head rotation transforms stored as Quaternions. Coverage feature is simply the number of unit cubes covered (area) by the user, calculated per frame. We used R and SPSS together to conduct the analysis of the trajectory data.

\section{Machine Learning Approach to Privacy - An LSTM Framework}
Deep learning is a dynamic tool to manipulate through data. Creating models in Deep Learning (DL) with Recurrent Neural Networks (RNN) would be one of the best fits with sequential data. In a maze-solving gaming paradigm, people generally tend to use previous memories to make instantaneous spatial decisions. Both Video gamers and non-video gamers tend to solve virtual mazes with previous information they have. Like if someone enters a dead end, they may try to remember their previous turns in the maze and get back to a point from where they can take another turn while avoiding the dead end. The strategy lies in the methodology of having memories of different states at different time periods. Long Short-Time Memory (LSTM) mimics a similar framework of reasoning. LSTM is a special type of RNN that traditionally contains cycles that feed the network activation from a previous time step as inputs to influence predictions at the current one\cite{r2}. The output of an LSTM network depends on the type of data. Depending on the problem, the output of each time step will have a different depiction. In robotics, the LSTM network has been implemented to demonstrate complex data and was able to verify the pattern of variable motion, which is analogous to human motion. 
The challenges of predicting human actions or the high degrees of freedom (DOF) human motions include inaccuracy and imperfections when they are estimated from real-time sequences\cite{doi:10.1177/0278364918812981}.
LSTM leverages the crucial ability to understand the context of sequential data. Independent of data size, LSTM can evaluate the prediction by learning to update and estimate the relative weights that represent the internal states through replication of its layers\cite{goodfellow2016deep}. Conventional RNNs have limited contextual information as the recurrent connections cause the input's influence to change exponentially, referred to as \textit{the vanishing gradient problem}. The LSTM type RNN is also no different in this regard, but a self-connected layer consisting of sub-networks with multiple internal cells leverages to get out of this problem. Multiplicative gates ease the way of storing and accessing information over long periods of time. This type of network helps to create an algorithm to take decisions either as a part of the instantaneous memory that should be removed or carried further during processing the sequential data\cite{Drumond2018PEEKA}.

\begin{figure}[tb]
 \centering 
 \includegraphics[width=\columnwidth]{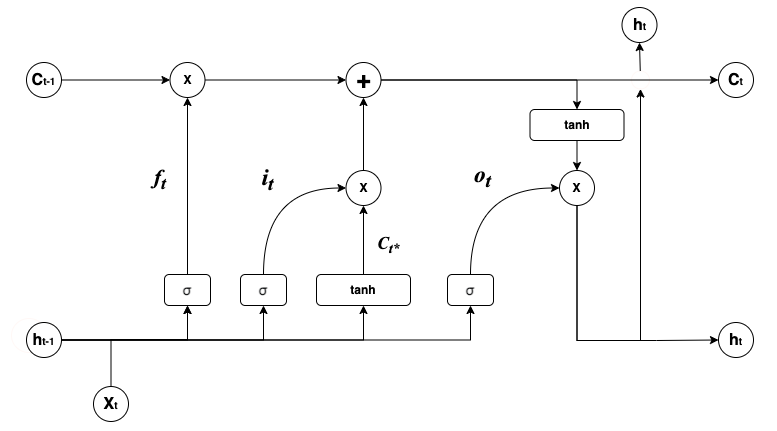}
 \caption{Schematic of LSTM single cell. $C_{t-1}$ represents the previous $t-1$-th step Cell State, $h_{t-1}$ the output of previous $t-1$-th step. $x_t$ is the $t$-th sample of the sequence, $C_t$ and $h_t$ are updated cell state and output respectively. $i_t$ , $o_t$,$f_t$ represents input, output, and forget gate respectively\cite{45500}.}
 \label{fig:lstmarchitecture}
\end{figure}

LSTMs can learn through data over a time period (much like the human brain) by controlling these gates, namely the input gate, forget gate, and output gate. These gates work as the write, read and reset of the memory cells. Memory cells are interconnected in an LSTM network in such a way that the output of one state can work as an input for the next cell. A schematic architecture of the LSTM network can be seen in \autoref{fig:lstmarchitecture}. This simplistic schematic shows, \textit{t}-th cell of the LSTM network, which receives the previous cell state $C_{t-1}$ representing the information being possessed so far until the previous \textit{(t-1)}-th step. Each LSTM cell receives input $x_t$ as the \textit{t}-th sample of a sequence along with previous states' output $h_{t-1}$ and processes an output value $h_t$ and updated cell state $C_t$. Gates of this LSTM network allow to make correlations during calculation and help information to be `forgotten' or `propagated' into the next cells \cite{goodfellow2016deep}.

An input gate $i_t$ and output gate $o_t$ control the input and output information respectively:
\[ i_t =\sigma( W_i. [h_{t-1},x_t] + b_i); \textit{[input gate]}\] 
\[ o_t =\sigma( W_o. [h_{t-1},x_t] + b_o); \textit{[output gate]}\]

\textit{W and b represent the weights and biases of the parameters, through which the information flow is controlled. }

New \textit{t}-th values will be updated using these equations:
\[ C_t = f_t*C_{t_1} + i_t*C_{t_*}\]
\textit{$C_{t_*}$ = computed state from the current step $x_t$ and previous output $h_{t-1}$}
\[ C_{t_*} = tanh(W_C.[h_{t_1},x_t]+b_C)\]
The final output value of a step is represented by this equation:
\[ h_{t_*} = o_t*tanh(C_t)\]

The forget gate $f_t$ helps the network to forget information that is no longer necessary: 
\[ f_t =\sigma( W_f. [h_{t-1},x_t] + b_f)\]

\section{Privacy Risks}
The usability of VR can broaden the capabilities for various activities; however, it also increases the potential for risks\cite{nla.cat-vn4854164}. Risk management in VR still lacks the user's trust in different perspectives such as privacy, security, well-being, etc. \cite{10.5555/3291228.3291263}. Similar concerns are being explored in the field of AR \cite{10.1145/2556288.2557352}. The question of privacy and security measures has still been left unanswered, nonetheless. All classes of users (including gamers and non-gamers) who opt to get an engaging experience using VR and AR platforms might have no idea about the amount of information they are giving away. Information gathered through VR or AR is a potential resource for imitating motor, visual or emotional patterns toward specific tasks. For instance, an individual can employ the deep face verification procedure without the assistance of a trained artificial intelligence model by using actual data obtained from a person's VR experience\cite{6909616}. 

The industry has started using `Synthetic Data', which can be produced using simplified data, which is not actual data. However, it can still produce mathematical patterns that are similar to those of real data\cite{SyntheticdataMITMSS}. Synthetic data's purpose is to safeguard privacy, but it can also be employed for other purposes. One of the most powerful tools available today, machine learning, can readily extract very accurate projected data from this type of input. Apple Inc. verifies FaceID using TrueDepth. To unlock a phone, TrueDepth employs attention-aware technology\cite{unlock}. FaceID detects information when someone looks at the smartphone with their eyes open and their attention focused there. As a result, it is more challenging for someone to unlock another device without permission. However, if any information is utilized to mimic that attention pattern, it may be possible to unlock it using that knowledge. This will probably be the case in the long run. 
Similarly, the test case of an immersive maze-solving video game can serve enough data to predict future steps with viable efficiency.

Numerous strategies are being suggested to mitigate privacy problems in the Metaverse. The user can create an addition to the clone of his avatar, allowing him to displace and change his appearance. An avatar can teleport to generate many copies of themselves that can be used to block tracking. Smart contracts should be built with countermeasures to companies trying to track users unethically.

Video game profiling is being researched to improve player performance and gaming experience. However, the gamer's privacy may be compromised if their gaming profile makes some internal information public \cite{Guardiola2015}. The viewpoint of game developers may be more favorable, but if exploited, it could be worse.

\section{Future work}
Privacy on the internet is already a significant concern and is not being adequately addressed \cite{5403147}; moreover, security concerns regarding VR applications have increased considerably. Spending a few hours in VR environments can provide developers with training data to create algorithms for pairing body attributes with subsequent behavior. The code of ethics can be modified for VR\cite{10.5555/3291228.3291263}, or users' permissions can be increased regarding sharing the data, but still, algorithms need to be implemented more generically. As stated earlier, ML is a powerful technology that can be applied to create algorithms in a preferred manner. ML should operate delicately, possibly separating the exploitative data from the usability data. Once ML has identified a pattern in the derived data, it can ask the users for permission to maintain or delete the pattern. It is equivalent to asking for consent before any medical procedure, regardless of the patient's desire to donate their body parts after death.

\section{Conclusions}
Many of the problems smartphones have are now passed onto AR/VR wearables intended for common users. One of the key issues these gadgets have in common is the ease with which they can take images, record videos, and track other potential usability data without proper vetting. The advent of smartphone cameras raised worries about third parties involved in the process, and there have been privacy worries about how simple it is to film bystanders. Due to the novelty of VR/AR display devices and the general lack of a standard of ethics-based practices, it is imperative that ML-driven approaches become sensitive to usability data to further reinforce trust and promote healthy usage of the technology.



\normalsize
\bibliography{references}


\end{document}